\documentclass[letterpaper, 10 pt, conference]{ieeeconf} 
\IEEEoverridecommandlockouts                                                                                   
\usepackage{amsmath,amsfonts}
\usepackage{graphics} 
\usepackage{epsfig} 
\usepackage{mathptmx} 
\usepackage{times} 
\usepackage{amssymb}  
\usepackage{algorithm}
\usepackage{algorithmic}
\usepackage{array}
\usepackage{textcomp}
\usepackage{stfloats}
\usepackage{url}
\usepackage{verbatim}
\usepackage{graphicx}
\usepackage{cite}
\usepackage{lipsum}
\usepackage{mathrsfs}  
\usepackage{mathtools}
\usepackage{braket}
\usepackage{multicol}
\usepackage{multirow}
\usepackage{float} 
\usepackage{booktabs} 
\usepackage{hyperref}
\graphicspath{{figs/}}
\title{\LARGE \bf
Real-Time Reconfiguration and Connectivity
Maintenance for AUVs Network Under External
Disturbances using Distributed Nonlinear Model Predictive Control
}
\author{Minh Nhat Nguyen$^{1,*}$, Stephen McIlvanna$^{1,*}$, Jack Close$^{1,*}$ and Mien Van$^{1}$ 
\thanks{$^{1}$Nhat Minh Nguyen, Stephen McIlvanna, Jack Close, and Mien Van (corresponding author) are with the School of Electronics, Electrical Engineering, and Computer Science, Queen’s University Belfast, Belfast, United Kingdom {\tt\small smcilvanna01@qub.ac.uk; nnhat01@qub.ac.uk; jclose06@qub.ac.uk; m.van@qub.ac.uk}}
\thanks{$^{*}$These authors contributed equally to this work.}}
\begin{document}
\maketitle
\thispagestyle{empty}
\pagestyle{empty}
\begin{abstract}
Advancements in underwater vehicle technology have significantly expanded the potential scope for deploying autonomous or remotely operated underwater vehicles in novel practical applications. However, the efficiency and maneuverability of these vehicles remain critical challenges, particularly in the dynamic aquatic environment. In this work, we propose a novel control scheme for creating multi-agent distributed formation control with limited communication between individual agents. In addition, the formation of the multi-agent can be reconfigured in real-time and the network connectivity can be maintained. The proposed use case for this scheme includes creating underwater mobile communication networks that can adapt to environmental or network conditions to maintain the quality of communication links for long-range exploration, seabed monitoring, or underwater infrastructure inspection. This work introduces a novel Distributed Nonlinear Model Predictive Control (DNMPC) strategy, integrating Control Lyapunov Functions (CLF) and Control Barrier Functions (CBF) with a relaxed decay rate, specifically tailored for 6-DOF underwater robotics. The effectiveness of our proposed DNMPC scheme was demonstrated through rigorous MATLAB simulations for trajectory tracking and formation reconfiguration in a dynamic environment. Our findings, supported by tests conducted using Software In The Loop (SITL) simulation, confirm the approach's applicability in real-time scenarios.
\end{abstract}
\section{INTRODUCTION}
The advancements in robotic technology, including communication elements, actuators and sensors, have allowed individual robotic agents to reach ever-rising levels of efficacy at many tasks. As with many human endeavours, task efficiency can potentially be further improved through the collaboration of multiple well-organised and directed agents. A wide range of control protocols have been investigated to improve the autonomy of multi-robot systems \cite{r1.1}. Key approaches in control design focus on decentralised feedback regulation \cite{r1.2}, fixed-time consensus fuzzy control \cite{r1.3}, and sliding mode control \cite{r1.4}. However, these approaches do not comprehensively consider aspects of performance optimality, including minimal control effort and travel distance. Furthermore, they are deficient in predicting future system outputs, handling state and control constraints, and executing replanning in real-time. In contrast, MPC emerges as a prominent method known for its systematic handling of system constraints and optimised control performance. Different MPC schemes have been proposed for trajectory tracking of a single Autonomous Underwater Vehicle (AUV) \cite{r1.5}-\cite{r1.7}, and multiple AUV systems \cite{r1.8},\cite{r1.9}.
\\
The challenges multiply when the focus shifts from individual to collective behaviour of robot systems. In the domain of multiple robot collaboration, \cite{r1.10} has highlighted the use of DMPC as a solution for enhancing the collaborative dynamics of mobile robots where each robot, by solving a locally constrained Optimal Control  Problem (OCP), achieves optimal control based on its own and neighbouring information. Moreover, DMPC strategies have been applied successfully for formation tracking and collision avoidance in groups of nonholonomic mobile robots, as discussed in \cite{r1.11}. Additionally, formation tracking methodologies for multi-AUV systems have been proposed using DMPC in \cite{r1.8},\cite{r1.9}. Despite the effectiveness of tracking and formation control presented in \cite{r1.5} -\cite{r1.9}, these approaches primarily focus on the kinematic properties or use simplified dynamic models of AUVs. Such limitations can render these schemes less effective in real-world applications. Furthermore, employing a simplified model may restrict the robots' ability to achieve comprehensive state awareness and hinder their collaborative potential.
\\
A centralised approach for a team of AUVs was also examined in \cite{r1.12}, utilising a global organiser and a local organiser to manage the team and distribute control commands to ensure coordinated behaviour amongst the AUVs. Similarly to \cite{r1.8},\cite{r1.9}, the proposed strategies necessitate each robot to communicate with the entire team or all followers linked to a leader. In these topological structures, the communication range between robots primarily determines the operational coverage of the team. They are not suitable for tasks that are carried out over a large area due to the limitations of current underwater wireless communication technologies. 
\\
As all agents must autonomously operate in dynamic environments with varying references and obstacles, safety and stability are important factors. CBF approach, as outlined in \cite{r2.6}, provides a structured framework extensively used to implement safety constraints like collision avoidance in robotic systems. \cite{r2.9} has proposed a new CBF with a decay-rate relaxation technique that can enhance both feasibility and safety. On the other hand, \cite{r2.9} demonstrated the improved performance and computational time of MPC with constraint-based CLF on a robot platform. 
\\
Motivated by the considerations outlined above, this work proposes a Distributed Nonlinear Model Predictive Control scheme (DNMPC), integrating CLF and CBF with a relaxed decay rate across the prediction horizon to the domain of AUVs (or underwater robotics) for the first time. Unlike \cite{r1.5} -\cite{r1.9}, the control scheme incorporates a 6-DOF dynamic model of AUV (i.e., BlueRov2), affording full spatial manoeuvrability and precise orientation control for executing complex collaborative tasks. Moreover, the distributed formation algorithm uses a network of AUVs as relaying units to maximise the team's operational coverage significantly. The effectiveness of the new control algorithm will be tested through complicated trajectory tracking and formation control tasks involving three AUVs in a dynamic environment with both static and dynamic obstacles. Additionally, a SITL simulation for each instance of our AUV (i.e., BlueRov2) system is also constructed with a robot operating system (ROS) network to emulate the communication structure.
\section{DYNAMIC MODELLING OF AUV}
\begin{figure}[h!]
\centering
\includegraphics[width=5cm]{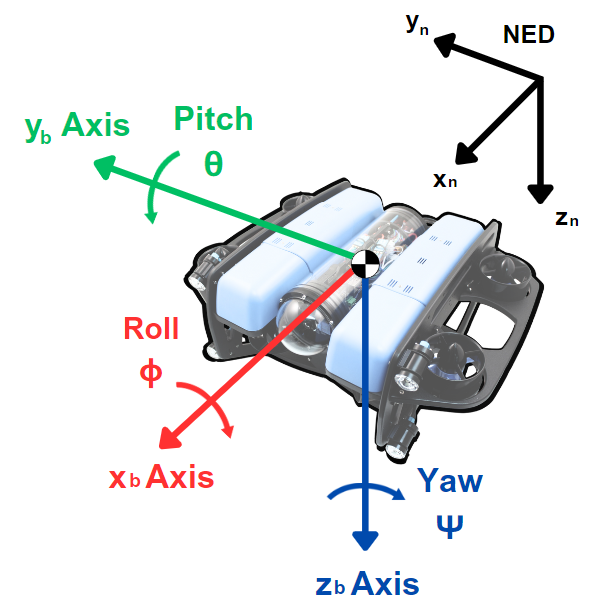}
\caption{BlueRov2 Degrees of Freedom.}
\label{DOF}
\end{figure}

We studied the coordinated motion control of a group of $n$ AUV (i.e., BlueRov2). For each AUV $i \in \{1,...,n\}$, the nomenclature prescribed to the AUV's 6-DOF variables comes from the Society of Naval Architects and Marine Engineers \cite{r2.1}. The reference frame of the environment is configured as North-East-Down (NED) with its centre denoted by $(x_{ni},y_{ni},z_{ni})$, which can be seen in Fig. \ref{DOF}.
\\
In the $i$-th AUV, the pose vector is defined as $\eta_i =[x_{bi},y_{bi},z_{bi},\phi_i,\theta_i,\psi_i]^T\in \Re^{6\times1}$, which includes the spatial coordinates of it's body centre $\eta_{pi} = [x_{bi},y_{bi},z_{bi}]^T$ (m) and the orientation angles $\eta_{oi} = [\phi_i,\theta_i,\psi_i]^T$ (rad) around the axes. Meanwhile, the velocity of the $i$-th AUV is defined by the velocity vector, $\mathcal{V}_i =[u_i,v_i,w_i,p_i,q_i,r_i]^T\in \Re^{6\times1}$. The directional velocities $\mathcal{V}_{di} = [u_i,v_i,w_i]^T$ (m/s) and the angular velocities $\mathcal{V}_{ai} = [p_i,q_i,r_i]^T$ (rad/s) represent the rate of turn around the $x$, $y$, and $z$ axes. The total state of the $i$-th AUV is denoted as $\xi_i = [\eta_i^T, \mathcal{V}_i^T]^T \in \Re^{12\times1}$. Furthermore, the control inputs are described by a vector $\tau_i=[\tau_{xi},\tau_{yi},\tau_{zi},\tau_{\phi i},\tau_{\theta i},\tau_{\psi i}]^T\in \Re^{6\times1}$ where $(\tau_{xi},\tau_{yi},\tau_{zi})$ (N) and $(\tau_{\phi i},\tau_{\theta i},\tau_{\psi i})$ (Nm) are control forces and torques for the $i$-th AUV, respectively. The description of the inertia matrix $M_i \in \Re^{6\times6}$, the Coriolis and centripetal matrix $C_i(\mathcal{V}_i) \in \Re^{6\times6}$, the hydrodynamic matrix $D_i(\mathcal{V}_i) \in \Re^{6\times6}$, the Jacobian matrix $J_i (\eta_i) \in \Re^{6\times6}$, and the vector of restoring forces $g_i(\eta_i) \in \Re^{6\times1}$ can be found in \cite{r2.2}. 
\\
This study addresses two core control challenges: stabilisation and trajectory tracking. For stabilisation, the goal is a constant target position with $\eta_r$ as the reference pose and a zero control vector $\mathcal{V}_r = 0^{6\times1}$. In trajectory tracking, both $\eta_r$ and $\mathcal{V}_r$ vary over time following a predetermined path. Both scenarios utilise a discretised dynamic model with a sampling period $\Delta t>0$. Denoting $\xi(k)=\xi(t_k)$, the integration over the fixed interval is numerically approximated with a piecewise constant control during each sampling interval and the Runge-Kutta 4 (RK4) method.
\begin{equation} \label{rk4}
\begin{aligned}
\xi_i(k+1) & = \textbf{F}(\xi_i(k),\tau_i(k))\\
& = \xi_i(k) + \int_{t_k}^{t_k + \Delta t} \mathbf{f}_i(\xi_i(t),\tau_i(t)) \,dt
\end{aligned}
\end{equation}
where $
\mathbf{f}_i=\left[ \begin{array}{c}
	J_i(\eta_i) \mathcal{V}_i\\
	M_i^{-1}(\tau_i -C_i(\mathcal{V}_i)\mathcal{V}_i-D_i(\mathcal{V}_i)\mathcal{V}_i-g_i(\eta_i))\\
\end{array} \right] .
$
\\
Denoting $k\in \mathbb{N}_0$ is the current sampling instant and $\mathbf{f}_i\in\Re^{12\times1}$ is the discrete nonlinear dynamic mapping of $i$-th AUV. In the stabilisation control problem, the feedback control is designed such that the solution of (\ref{rk4}) starting from the initial condition $\xi_{0,i}:=\xi_i(0)\in \mathbf{X}$ stays close to a desired set point, $\xi _{r,i}\in \mathbf{X}$, and converges, i.e.
\begin{equation} \label{e1}
\underset{k\rightarrow \infty}{\lim}\|\xi_{e,i}\|=\underset{k\rightarrow \infty}{\lim}\| \xi_i(k) - \xi_{r,i}\|=0
\end{equation}
For the trajectory control problem, the feedback control task is to steer the solution of (\ref{rk4}) to track the time-varying reference such that
\begin{equation} \label{e2}
\underset{k\rightarrow \infty}{\lim}\|\xi_{e,i}\|=\underset{k\rightarrow \infty}{\lim}\| \xi_i(k) - \xi_{r,i}(k)\|=0
\end{equation}
\section{Control Design}
\begin{figure}[h!]
\centering
\includegraphics[width=7cm]{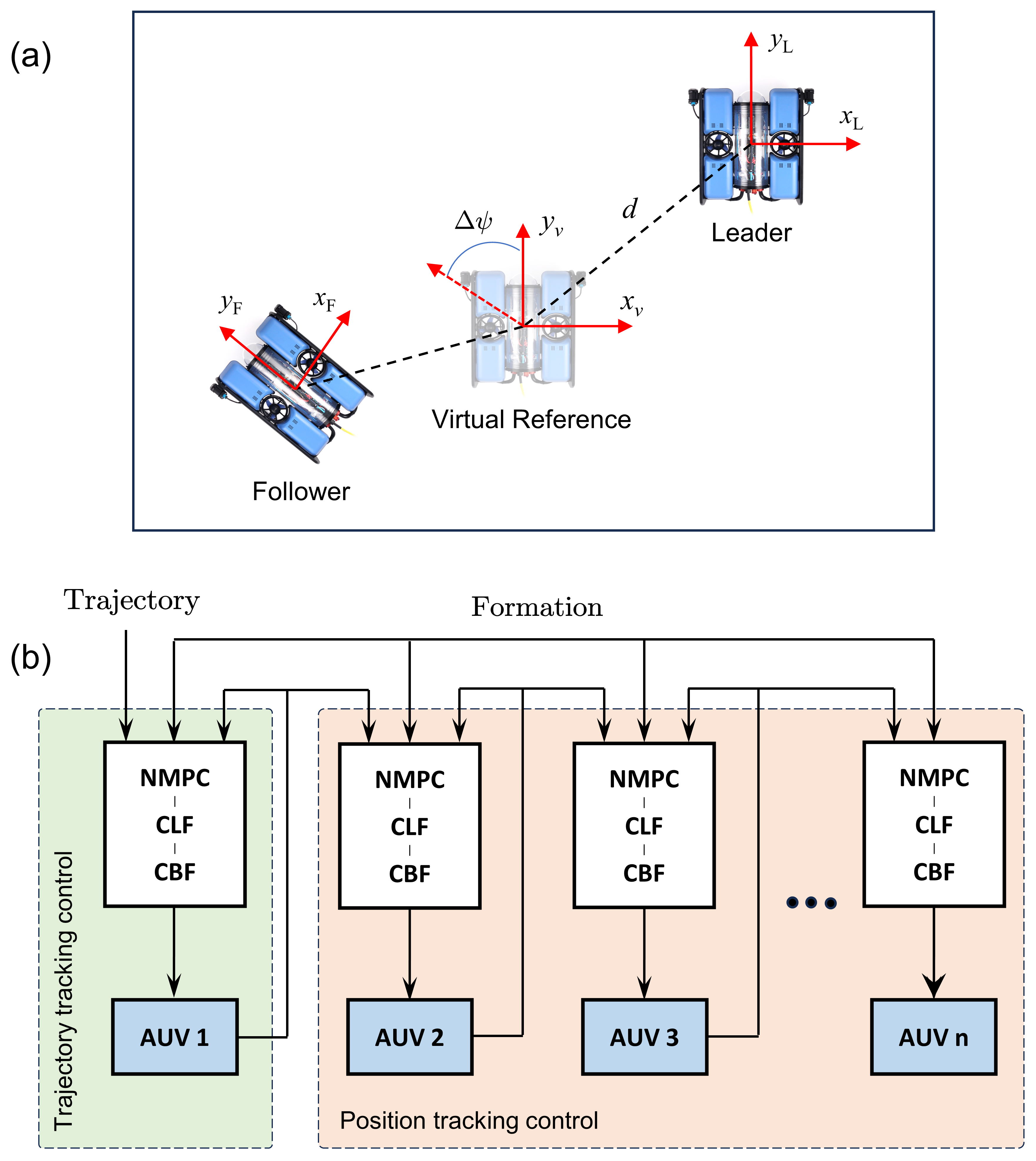}
\caption{A pair of leader-follower formation.}
\label{pair}
\end{figure}

Consider $n$ AUV agents operating in a workspace $W \subset \Re^3$. Within the formation, there are numerous pairs of leaders and followers, and the team of $n$ agents can be decomposed into $n-1$ decentralised subsystems, each comprising two AUVs. In each subsystem, the follower AUV strives to maintain a desired distance and angles relative to its leader. The desired formations are established when all vehicles reach their expected positions. As shown in Fig. \ref{pair} (a), we formulate one pair leader-follower formation problem as follows: Given the pose $\eta_L$ of a leader vehicle, the reference trajectory for the follower is set such that its position is shifted by a distance $d$ and angles $(\Delta \phi,\Delta \theta,\Delta \psi)$ relative to the leader. The reference trajectory of the follower is generated as the leader cruises. This study employs a network of AUVs as relaying units to extend the operational coverage capacity of the team. In this setup, only AUV1 is directly informed of the desired trajectory by a base computer. Subsequently, AUV1 broadcasts a set of information containing its current state, denoted as $\xi_{1}$, and a formation vector, $\mathcal{A}=[\Delta x,\Delta y,\Delta z,\Delta \phi,\Delta \theta,\Delta \psi]^T$, to its follower. Similarly, AUV3 receives the same set of information from AUV2, as demonstrated in Fig. \ref{pair} (b).
\subsection{Proximity Graph}
The connectivity of the robot team through communication is represented by a proximity graph, $\mathcal{G} = (H, E(t))$, where $H$ is the vertices set corresponding to robots $i \in {1, \ldots, n}$, and $E(t)$ is the edges set indicating communication links at the time step $t$. A link between AUV $i$ and AUV $j$ exists if their Euclidean distance does not exceed a specified communication range, $d_c$. Therefore, $E(t)$ is determined by
\begin{equation} \label{eq1}
E(t)=\{(i,j)\,|\,\, \|\eta_{p,i}(t)-\eta_{p,j}(t)\| \leqslant d_c, \forall{i,j}\in V, i\ne j\}
\end{equation}
A pair of AUVs $(i,j)$ are said to be neighbours at time step $t$ if $(i,j)\in E(t)$ and the set of neighbours of each AUV $i$ is defined as $\mathcal{N}_i (t)=\{j| (i,j)\in E(t)\}$. It is worth noting that the proximity graph is dynamic as both $E(t)$ and $\mathcal{N}_i(t)$ are time-varying when the AUVs move around.
%
\subsection{Control Lyapunov Function}
This section presents the Control Lyapunov Function (CLF) that enforces the stability of the nonlinear discrete-time control system (\ref{rk4}). Some necessary definitions are introduced below.
\\
$\textbf{Definition 1.}$ A function $\alpha \,\,: \Re_{\geqslant 0}\rightarrow \Re_{\geqslant 0}$ is a class $\mathcal{K}$-function if it is continuous, strictly increasing and $\alpha(0)=0$. A function $\alpha \,\,:\Re_{\geqslant 0}\rightarrow \Re_{\geqslant 0}$ is of class $\mathcal{K}_{\infty}$-function if it is a $\mathcal{K}$-function and $\alpha(s)\rightarrow +\infty$ as $s \rightarrow +\infty$.
\\
$\textbf{Definition 2.}$ An admissible control invariant set $\mathbf{X}$ for system (\ref{rk4}) is defined as a set where, for all $\xi(k) \in \mathbf{X}$, there exists a state-feedback controller $\tau(k) \in \mathbf{U}$ such that $\mathbf{F}(\xi(k), \tau(k))\in \mathbf{X}$. Furthermore, a function $V$ is considered a CLF in $\mathbf{X}$ if, for all $\xi(k) \in \mathbf{X}$, there exist three $\mathcal{K}_{\infty}$-functions $\alpha_1$, $\alpha_2$, and $\alpha_3$ that satisfy $\alpha_1(\|\xi_e(k)\|) \leqslant V(\xi_e(k)) \leqslant \alpha_2(\|\xi_e(k)\|)$ and: 
\begin{equation}\label{eq2}
\varDelta V(\xi_e(k),\tau(k)) +\alpha_3(\|\xi_e(k)\|) \leqslant 0
\end{equation}
By using $\alpha_1$, $\alpha_2$, $\alpha_3$ $\in \mathcal{K}$-function, the upper bound on the CLF is essential for ensuring Lyapunov stability guarantees for any horizon length $N$ of NMPC. These CLF constraints establish the existence of a point-wise set of control inputs that achieve stabilisation:
\begin{equation}\label{eq3}
\kappa_{clf}=\{\tau \in \mathbf{U}: \varDelta V(\xi_e(k),\tau(k))\leqslant - \alpha V(\xi_e(k))\}
\end{equation}
where $\varDelta V(\xi_e(k),\tau(k)) = V(\xi_e(k+1))-V(\xi_e(k))$
$\textbf{Remark 1.}$ Refer to \cite{r2.3} for a comprehensive stability analysis of the CLF constraint integrated with NMPC. Furthermore, the research in \cite{r2.4} demonstrates that this method has several desirable properties, such as the absence of a terminal cost and the requirement for fewer tuning parameters. Additionally, \cite{r2.5} successfully applied NMPC with the CLF constraint on a robot platform, resulting in improved computational efficiency.
%
\subsection{Control Barrier Function}
\begin{figure}[h!]
\centering
\includegraphics[width=7cm]{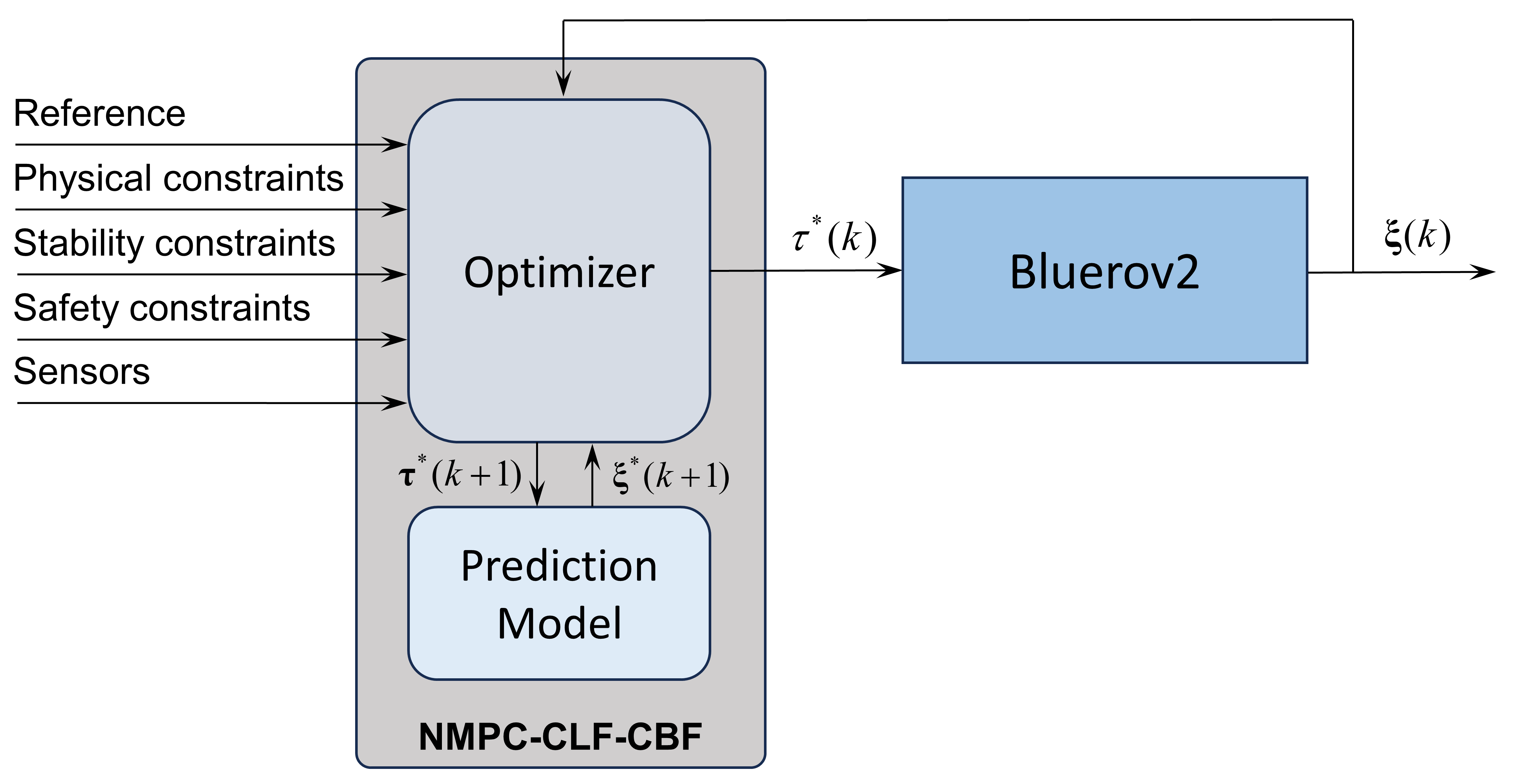}
\caption{Diagram of the proposed framework.}
\label{fig1}
\end{figure}

This section derives a safety constraint for the optimal control input that guarantees the robot's spatial position, $\eta_p$, always lies within a defined safe set, $\mathscr{Z} \in \Re^{3}$.
\begin{subequations}\label{eq4}
\begin{align}
&\label{eq18a}\mathscr{Z} =\{\eta_{p} \in \mathbf{X}\subset \Re^3|B(\eta_{p}) \geqslant 0 \},\\
&\label{eq18b}\partial \mathscr{Z} =\{\eta_{p} \in \mathbf{X}\subset \Re^3|B(\eta_{p}) =0 \},\\
&\label{eq18c}\mathrm{Int}(\mathscr{Z}) \{\eta_{p} \in \mathbf{X}\subset \Re^3|B(\eta_{p}) >0 \}.
\end{align}
\end{subequations}
Let $\mathscr{Z}$ denote the superlevel set of a continuously differentiable function $B:\Re^3\rightarrow \Re$. The function $B$ qualifies as a CBF if there exists an $\mathcal{K} _{\infty}$-function $\gamma$ for the control system (\ref{rk4}) to satisfy:
\begin{equation} \label{eq5}
\exists \,\,\tau \,\,\mathrm{s.t.} \,\,\dot{B}(\eta_{p},\tau) \geqslant -\gamma B( \eta_{p}), \gamma \in \mathcal{K} _{\infty} 
\end{equation}
Extending this inequality constraint to the discrete-time domain and employing $\gamma$ as a scalar, the set comprising all control values at a given point $\eta_{p}(k) \in \mathbf{X}$:
\begin{equation} \label{cbf}
\begin{aligned}
\kappa_{cbf}(\eta_{p}(k)) =\{\tau \in \mathbf{U}&: \Delta B(\eta_{p}(k),\tau) +\gamma B( \eta_{p}(k)) \geqslant 0,\\  
&0\leqslant \gamma \leqslant 1\} 
\end{aligned}
\end{equation}
$\textbf{Remark 2.}$ By Theorem 2 in \cite{r2.6}, if $B$ is a CBF in $\mathbf{X}$ and $\frac{\partial B}{\partial \eta_{p}}(\eta_{p}) \ne 0$ for all $\eta_{p} \in \partial \mathscr{Z}$, then any control signal $\tau\in \kappa_{cbf}(\eta_{p}(k))$ for the system (\ref{rk4}) renders the set $\mathscr{Z}$ safe. Additionally, the set $\mathscr{Z}$ is asymptotically stable in $\mathbf{X}$.
\\
The rigid body obstacles are conceptualised as the union of spheres with centroids ($x_{ob,i}, y_{ob,i}, z_{ob,i}$) and a fixed radius $r_{ob,i}$. Similarly, the AUV safety zone is defined by a sphere centred on the AUV with a radius $r_{rb}$. Safety sets for obstacles are denoted as follows:
\begin{equation} \label{eq7}
\mathscr{Z} _i=\{ \eta_{p}(k) \in \mathbf{X}\subset \Re^3:B_i( \eta_{p}(k),\mathcal{P}(k)) \geqslant 0 \}
\end{equation}
where 
\begin{equation} \label{eq8}
\begin{aligned}
B_i(\eta_{p}(k),\mathcal{P}(k)) =&(x_b(k)-x_{ob,i}(k))^2+(y_b(k) -y_{ob,i}(k))^2\\&
+(z_b(k)-z_{ob,i}(k))^2-(r_{rb}+r_{ob,i})^2
\end{aligned}
\end{equation}
Assuming that the AUV can detect the position and size of obstacles, $\varphi_{ob}=(x_{ob},y_{ob},z_{ob},r_{ob})^T$, through its sensors at each sampling time, this information is then stored inside the parameter vector $\mathcal{P}$. To ensure safety while regulating the target state, the conditions in (\ref{cbf}) are added to the NMPC as safety constraints.
By incorporating the CLF and CBF constraints into the proposed NMPC schemes, the team of agents can safely follow the desired path $\xi_r(t)$ while guaranteeing the maintenance of the predefined formation geometry and the connectivity among the agents. The control scheme is depicted in Fig. \ref{fig1}. 
%
\subsection{NMPC-CLF-CBF for Trajectory Tracking Control}
The following section defines the controller for the trajectory tracking control of AUV1 as it is set up to have access to the desired trajectory. The conventional NMPC scheme is formulated by constructing and minimising a cost function and constraints on the prediction horizon grids $N$. The OCP is presented in parametric form as follows: 
\begin{subequations}\label{ocp1}
\begin{align}
&\begin{aligned}\mathbf{OCP}_1\,:\,& J_{1N}(\xi_1 (k),\tau_1(k),\mathcal{P}_1(k)) = \underset{{\mathbf{S}_1}}{\min}\underset{m=0}{\overset{N-1}{\sum}}(\varLambda_1(s_v(m))\\ &+ \varGamma_1(s_h(m))+\ell_1( \xi_1(k+m), \tau_1(k+m), \mathcal{P}_1(k))) \end{aligned}\\ 
&\text{Subject to} \nonumber \\ 
&\xi_1(k) - \xi_{1\text{fb}} = 0,\\
&\xi_1(k+m+1) - \mathbf{F}_1(\xi_1(k+m),\tau_1(k+m)) = 0,\\
&\xi_1(k+m) \in \mathbf{X},\,\,\tau_1(k+m) \in \mathbf{U},\\
&\begin{aligned}&(1-\alpha)V_1(\xi_1(k+m),\mathcal{P}_1(k))+ s_v(m)\\ &-V_1(\xi_1(k+m+1),\mathcal{P}_1(k)) \geqslant 0,\,\,\end{aligned}\\
&\begin{aligned}&B(\eta_{1p}(k+m+1),\mathcal{P}_1(k))\\ &- s_h(m)(1-\gamma)B(\eta_{1p}(k+m),\mathcal{P}_1(k)) \geqslant 0.\end{aligned}
\end{align}
\end{subequations}
where $s_v$ are relaxation terms that avoid conflict between CLF and CBF constraints, and $s_h$ are decay-rate slack variables added to enhance the feasibility of the OCP. Hence, the $\varLambda_1(s_v)$ and $\varGamma_1(s_h)$ are included in the cost function to minimise those slack variables. 
\begin{equation}\label{slack}
\begin{aligned}
&\varLambda_1(s_v(m)) = W_1s_v(m)^2,  \\
&\varGamma_1(s_h(m)) = W_2(s_h(m)-1)^2.
\end{aligned}
\end{equation}
\\
The stage cost includes two quadratic terms: one will penalise the tracking error and another will minimise the control outputs, $\ell_1:\Re^{12}\times \Re^6\rightarrow \Re_{\geqslant 0}$. The CLF is a quadratic function of the state variable error, $V_1:\Re^{12}\rightarrow \Re_{\geqslant 0}$.
\begin{equation}\label{eq9}
\begin{aligned}
&\begin{aligned}\ell_1(\xi_1(k),\tau_1(k),\mathcal{P}_1(k))=&(\xi_{1}(k)-\xi_{r}(k))^T W_{\xi 1} (\xi_{1}(k)-\xi_{r}(k))\\ &+\tau_1(k)^TW_{u1} \tau_1(k),\end{aligned}\\
&V_1(\xi_1(k),\mathcal{P}_1(k))=(\xi_{1}(k)-\xi_{r}(k))^TW_{v1}(\xi_{1}(k)-\xi_{r}(k)).
\end{aligned}
\end{equation}
where $W_{\xi1}$, $W_{u1}$, and $W_{v1}$ are positive definite symmetric matrices. The decision variables are $\mathbf{S}_1=[\xi_1(k)^{T},...,\xi_1(k+N)^{T},\tau_1(k)^{T},...,\tau_1(k+N-1)^{T}, s_v(k),..., s_v(k+N-1), s_h(k),...,s_h(k+N-1)]$.
\\
The multiple shooting approach \cite{r2.7} is used to transform the OCP into a Nonlinear Program (NLP). $\mathcal{P}_1$ is a problem parameter vector that contains reference state vector $\xi_r$, feedback state vector $\xi_{1\text{fb}}$, formation vector $\mathcal{A}$, and obstacles information. Constructing a vector that contains all decision variables as $\varXi_1 =\mathbf{S}_1^T$, the NLP is introduced as below:
\begin{equation} \label{nlp1}
\underset{\varXi_1}{\min}\,\,A_1\left( \varXi_1 ,\mathcal{P}_1 \right) \,\,s.t. \left\{ \begin{aligned}
	&H_1\left( \varXi_1 ,\mathcal{P}_1 \right) =0,\\
	&G_1\left( \varXi_1 ,\mathcal{P}_1 \right) \leqslant 0.
\end{aligned} \right. 
\end{equation}
The NLP constraints are divided into two categories: the equality constraint vector, denoted as $H_1(\varXi_1, \mathcal{P}_1)$, ensures that the system dynamics in (\ref{ocp1}b) and (\ref{ocp1}c) are met, and the inequality constraint vector, represented by $G_1(\varXi_1, \mathcal{P}_1)$, imposes restrictions of (\ref{ocp1}e) and safety constraints (\ref{ocp1}f) on the decision variables to maintain the system within safe boundaries.
%
\subsection{NMPC-CLF-CBF for Position Tracking Control}
The follower’s objective is to maintain a desired relative position with respect to the leader. All other agents $i = 2,...,n$ do not possess any knowledge of the desired trajectory. Within a pair of AUVs, the leader broadcasts its state and formation vector at time $k$ to its follower. Hence, the control framework for position tracking is proposed as follows: 
\begin{subequations}\label{ocp2}
\begin{align}
&\begin{aligned}\mathbf{OCP}_i\,:\,& J_{iN}(\xi_i (k),\tau_i(k),\mathcal{P}_i(k)) = \underset{{\mathbf{S}_i}}{\min}\underset{m=0}{\overset{N-1}{\sum}}(\varLambda_i(s_v(m))\\ &+ \varGamma_i(s_h(m))+\ell_i( \xi_i(k+m), \tau_i(k+m), \mathcal{P}_i(k))) \end{aligned}\\ 
&\text{Subject to} \nonumber \\ 
&\xi_i(k) - \xi_{i\text{fb}} = 0,\\
&\xi_i(k+m+1) - \mathbf{F}_i(\xi_i(k+m),\tau_i(k+m)) = 0,\\
&\xi_i(k+m) \in \mathbf{X},\,\,\tau_i(k+m) \in \mathbf{U},\\
&\begin{aligned}&(1-\alpha)V_i(\xi_i(k+m),\mathcal{P}_i(k))+ s_v(m)\\ &-V_i(\xi_i(k+m+1),\mathcal{P}_i(k)) \geqslant 0,\,\,\end{aligned}\\
&\begin{aligned}&B(\eta_{ip}(k+m+1),\mathcal{P}_i(k))\\ &- s_h(m)(1-\gamma)B(\eta_{ip}(k+m),\mathcal{P}_i(k)) \geqslant 0.\end{aligned}\\
&\mathcal{C}_i(\eta_{pL}, \eta_{pi}(k+m)) \leqslant d_c.
\end{align}
\end{subequations}
All constraints of OCP (\ref{ocp2}) are set up similarly to the OCP (\ref{ocp1}), only additional constraint (\ref{ocp2}g) guarantees the follower stays within the communication range with its leader. The stage cost, CLF, and distance constraint $\mathcal{C}_i$ are define as:
\begin{equation}\label{eq9}
\begin{aligned}
&\begin{aligned}\ell_i(\xi_i(k),&\tau_i(k),\mathcal{P}_i(k))=\tau_i(k)^TW_{ui} \tau_i(k)\\&+((\xi_{L}+\mathcal{A}_i)-\xi_{i}(k))^T W_{\xi i} ((\xi_{L}+\mathcal{A}_i)-\xi_{i}(k)),\end{aligned}\\
&V_i(\xi_i(k),\mathcal{P}_i(k))=((\xi_{L}+\mathcal{A}_i)-\xi_{i}(k))^TW_{vi}((\xi_{L}+\mathcal{A}_i)-\xi_{i}(k)),\\
&\mathcal{C}_i(\eta_{pL}, \eta_{pi}(k))=\sqrt{(x_{bL}-x_{bi})^2+(y_{bL}-y_{bi})^2+(z_{bL}-z_{bi})^2}.
\end{aligned}
\end{equation}
where $W_{\xi i}$, $W_{ui}$, and $W_{vi}$ are positive definite symmetric matrices. 
\\
Similar NLP for other agents in the team are constructed
\begin{equation} \label{nlp2}
\underset{\varXi_i}{\min}\,\,A_i\left( \varXi_i ,\mathcal{P}_i \right) \,\,s.t. \left\{ \begin{aligned}
	&H_i\left( \varXi_i ,\mathcal{P}_i \right) =0,\\
	&G_i\left( \varXi_i ,\mathcal{P}_i \right) \leqslant 0.
\end{aligned} \right. 
\end{equation}
\\
The NLP in (\ref{nlp1}) and (\ref{nlp2}) are solved using an Interior Point OPTimizer (IPOPT) by the local processor of each agent. Based on the current feedback state and environmental data, the IPOPT solver derives an optimal state and control sequence, $\boldsymbol{\xi }_{iN}^*$ and $\boldsymbol{\tau}_{iN}^*$. The first element of $\boldsymbol{\tau}_{iN}^*$, labelled $\tau^*_i(k)$, is then applied to the system. The residual part of the optimised sequence serves as an initial guess for the decision variable vector in the subsequent iteration. Thereafter, the solver is re-engaged to deduce new control input values. An outline of this control methodology is delineated in Algorithm 1.
\begin{algorithm}[H]
\caption{IPOPT - NMPC}\label{alg1}
\begin{algorithmic}
\STATE 
\STATE $ \textbf{Given}$  $\varXi_i^0$
\STATE $ \textbf{Initialise}$  $(m,\varXi^m_i) \gets (0,\varXi^0_i)$
\STATE $ \textbf{while}$  $\text{ControllerIsRunning()}$  $\textbf{do}$
\STATE \hspace{0.5cm}$\xi_{i\text{fb}}$ $\gets$ $\text{StateFeedback()}$
\STATE \hspace{0.5cm}$\xi_r$/$\xi_L$, $\mathcal{A}_i$ $\gets$ $\text{Commands()}$
\STATE \hspace{0.5cm}$\varphi_{ob,i}$ $\gets$ $\text{Sensors}$
\STATE \hspace{0.5cm}$\mathcal{P}_i$ $\gets$ $(\xi_{\text{fbi}},\xi_r/\xi_L,\mathcal{A}_i,\varphi _{ob,i})$
\STATE \hspace{0.5cm}$\varXi^{m+1}_i$ $\gets$ $\text{IPOPT solver}(\mathcal{P}_i,\varXi^m_i)$
\STATE \hspace{0.5cm}$\boldsymbol{\tau}_{iN}^*$ $\gets$ $\text{ExtractInputSequence}(\varXi^{m+1}_i)$
\STATE \hspace{0.5cm}$\tau^*_i(k)$ $\gets$ $\text{ExtractFirstInput}(\boldsymbol{\tau}_{iN}^*)$
\STATE \hspace{0.5cm}$\mathbf{f}(\xi,\tau)$ $\gets$ $\text{ApplyInput}(\tau^*_i(k))$
\STATE \hspace{0.5cm}$\varXi^m$ $\gets$ $\text{Shift}(\boldsymbol{\xi }_N^*,\boldsymbol{\tau}_{iN}^*)$
\STATE \hspace{0.5cm}$m$ $\gets$ $m+1$
\STATE $ \textbf{end while}$ 
\end{algorithmic}
\end{algorithm}
$\textbf{Remark 3.}$ As analysed in \cite{r2.8}, the proposed control schemes with relaxed CLF and CBF constraints consistently maintain optimisation feasibility. Furthermore, with the implementation of the decay-rate relaxation technique for CBF, as introduced by \cite{r2.9}, the safety performance of agents is enhanced by reducing $\gamma$ while not harming the feasibility of the OCP.
\section{Results}  
To verify our proposed frameworks, we considered the control of a team of three AUV (i.e., BlueRov2) systems that operated in a dynamic environment containing random static and moving obstacles. The whole team followed a desired trajectory $\xi_r = [0.7t-3,-3,2,0,0,0]^T$ from $0-70$ seconds, then, $\xi_r = [0.5t+11.3,-3,0.5t-32.7,0,0,0]^T$ from $71-110$ seconds. Three AUVs started from the initial positions: $\eta_1(0)=[-3,-3,-3,0,0,0]^T$, $\eta_2(0)=0_{6\times1}$, $\eta_3(0)=[0,-7,1,0,0,0]^T$, and $\mathcal{V}_i=0_{6\times 1}, i \in \{1,2,3\}$ was set for the initial velocities of AUVs. Moreover, the team formed four different formations, as shown in Table. 1.
\begin{table*}[h!]
    \centering
    \caption{Values of $\mathcal{A}_2$ and $\mathcal{A}_3$ for corresponding time intervals.}
    \label{table:A_values}
    \begin{tabular}{|c|c|c|}
    \hline
    Time Interval ($t$) & $\mathcal{A}_2$ & $\mathcal{A}_3$ \\ 
    \hline
    $0 \leqslant t \leqslant 30$ & $[0,1,4,0,0,0]^T$ & $[0,-2,0,0,0,0]^T$ \\
    $30 < t \leqslant 70$ & $[-5,0,0,0,0,0]^T$ & $[-5,0,0,0,0,0]^T$ \\
    $70 < t \leqslant 90$ & $[-0.56t+39.7, 0.07t-5,0,0,0,-0.08t+5.5]^T$ & $[0,-0.14t+9.9,0,0,0,0.16t-11]^T$ \\
    $90 < t \leqslant 110$ & $[-0.56t+39.7, 0.07t-5,0,0,0.1t-10.5,0]^T$ & $[0,-0.14t+9.9,0,0,0.1t-10.5,0]^T$ \\
    \hline
    \end{tabular}
\end{table*}
\begin{figure}[h!]
\centering
\includegraphics[width=8cm]{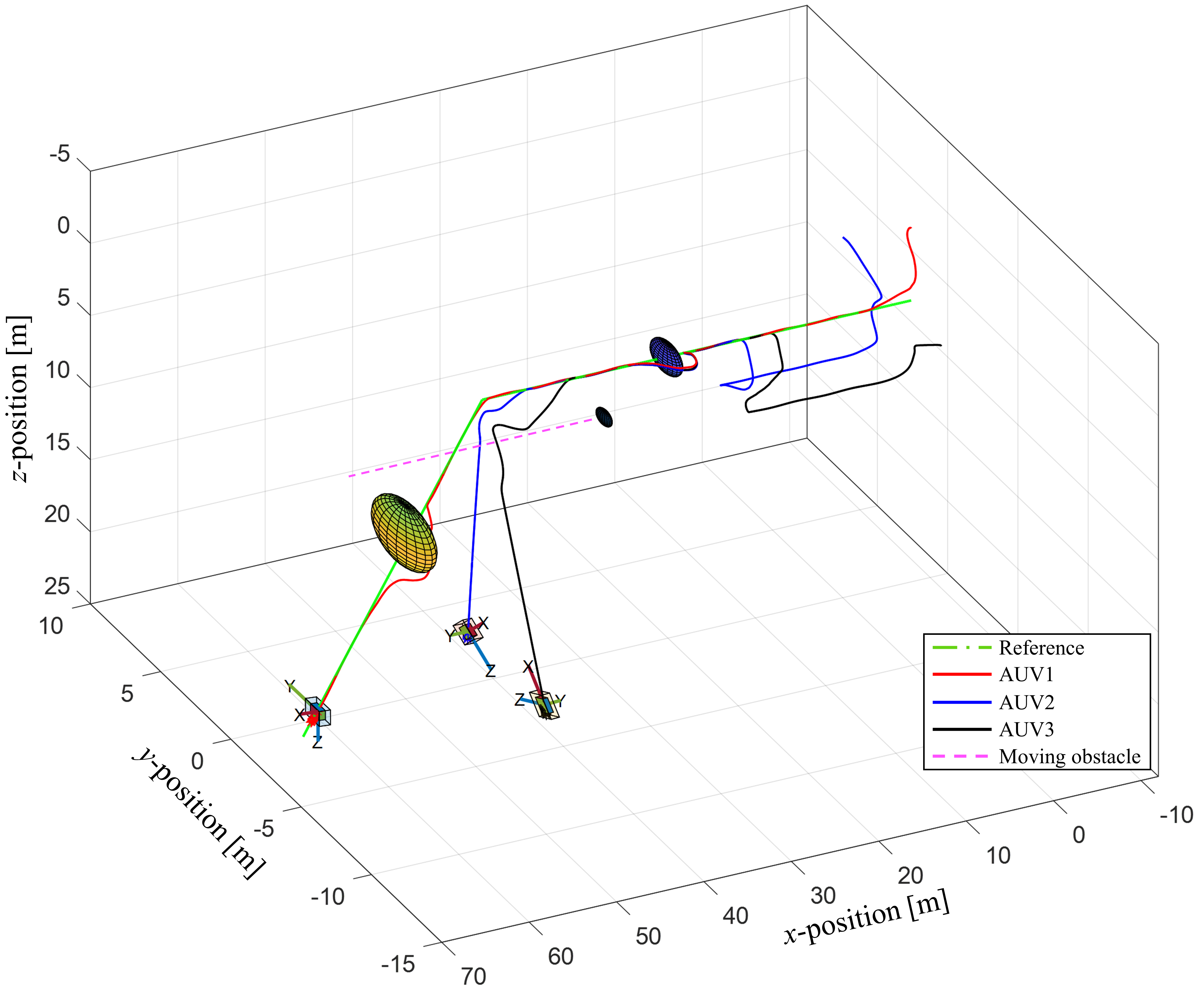}
\caption{Trajectory tracking and formation control in a dynamic environment.}
\label{trajectory}
\end{figure}
\begin{figure}[h!]
\centering
\includegraphics[width=8cm]{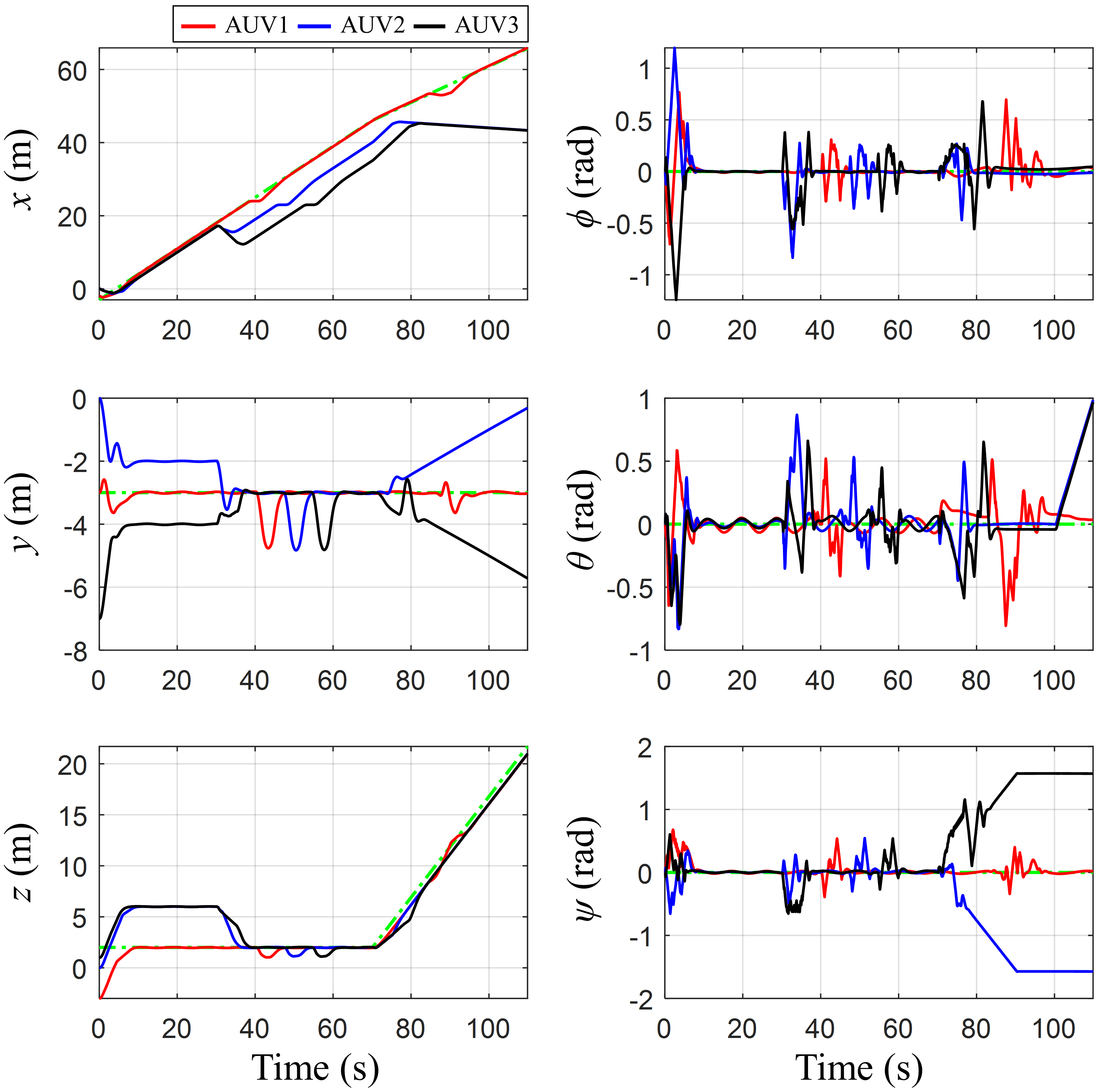}
\caption{Tracking performance of the system pose state.}
\label{state}
\end{figure}
During the tracking task, a challenging scenario was designed with two static obstacles placed across the reference path at (25m,-3m,2m) and (55m,-3m,10m), with the radius of each obstacle being 1 m and 2 m, respectively. Additionally, we introduced one moving obstacle to the scenario. The initial pose, velocity, and radius of the moving obstacles were given as (70,-4,4,-0.33,0,0,0.5)\,\,(m,m,m,m/s,m/s,m/s,m). The simulation is set up on MATLAB Simulink with an Intel\textsuperscript{\textregistered} $\text{Core}^{\text{TM}}$ i7-1185G7 and 16GB RAM. 
\begin{figure}[!htb]
\centering
\includegraphics[width=8cm]{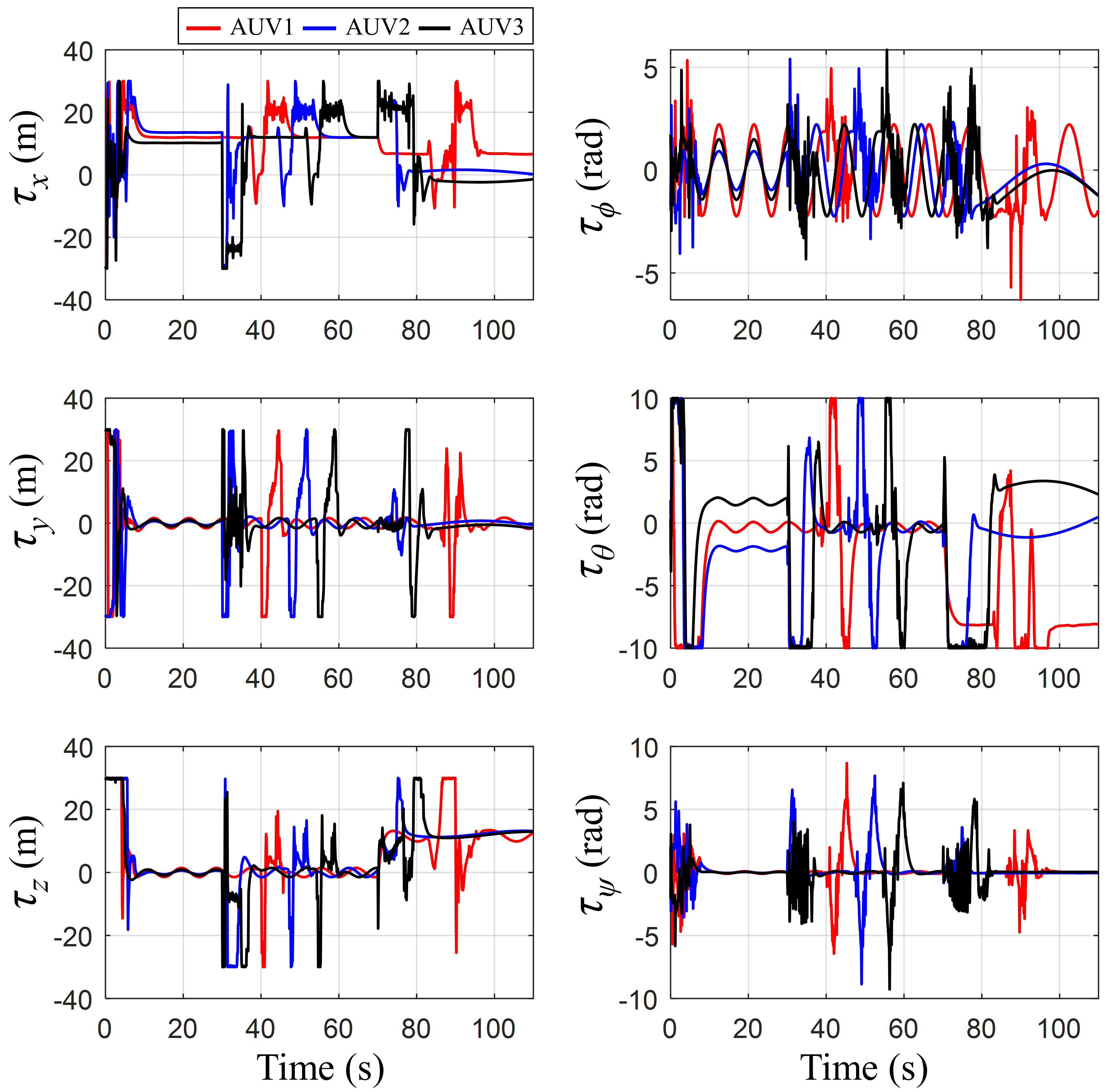}
\caption{Control signal of the proposed DNMPC.}
\label{control}
\end{figure}
\\
The overall trajectories of the proposed NMPC-CLF-CBF for three BlueRov2 in the 3D space are plotted in Fig. \ref{trajectory}. The pink line represents the trajectory of moving obstacles. All agents exhibited efficient tracking performance when robots did not encounter obstacles. Fig. \ref{state} illustrates the tracking performance of each AUV in the scenario, showcasing smooth, optimized movements from initial states to time-variant reference postures throughout the duration of the simulation. Upon examining the plots for $x, y,$ and $z$, it was noticeable that AUV1—the leader—accurately followed the reference trajectory and only deviated when encountering obstacles. Similarly, AUV2 and AUV3 maintained a stable formation relative to the leader and independently executed collision avoidance, irrespective of the leader’s movements. This independence was due to each AUV having its own distinct CBF constraint for collision avoidance, informed by feedback from its unique environmental interactions at each point in time. The plots for $\phi, \theta,$ and $\psi$ depicted the orientation of each AUV as they navigated and adapted within the three-dimensional space. The change in the yaw angle between 70-90 s and in the pitch angle between 100-110 s demonstrated that all states of the robots can be controlled, which can enhance the ability to perform complex collaborative tasks. All the control signals were within their limitations, and large oscillations were observed as the robots navigated around the obstacles, as shown in Fig. \ref{control}. Moreover, the effectiveness of the proposed controller was supported by tests conducted on multiple Bluerov2 models using SITL simulation, as presented in Fig. \ref{ros}. Three BlueRov2 AUVs followed a linear trajectory from 0 to 150 m with varying speeds along the path. Three different formations were configured during the tracking, as seen in Fig. \ref{ros} (b). The experimental results aligned with the outcomes obtained from the Matlab simulations, showing that both trajectory tracking and formation control were achieved. A video record the simulation results in detail is available online (See Video 1). Physical experiments are being carried out with three physical BlueRov2s in both water tanks and open sea (see our physical experimental setup: https://youtu.be/7752fxqaIss) and the results will be presented in our future works.
\begin{figure}[h!]
\centering
\includegraphics[width=7cm]{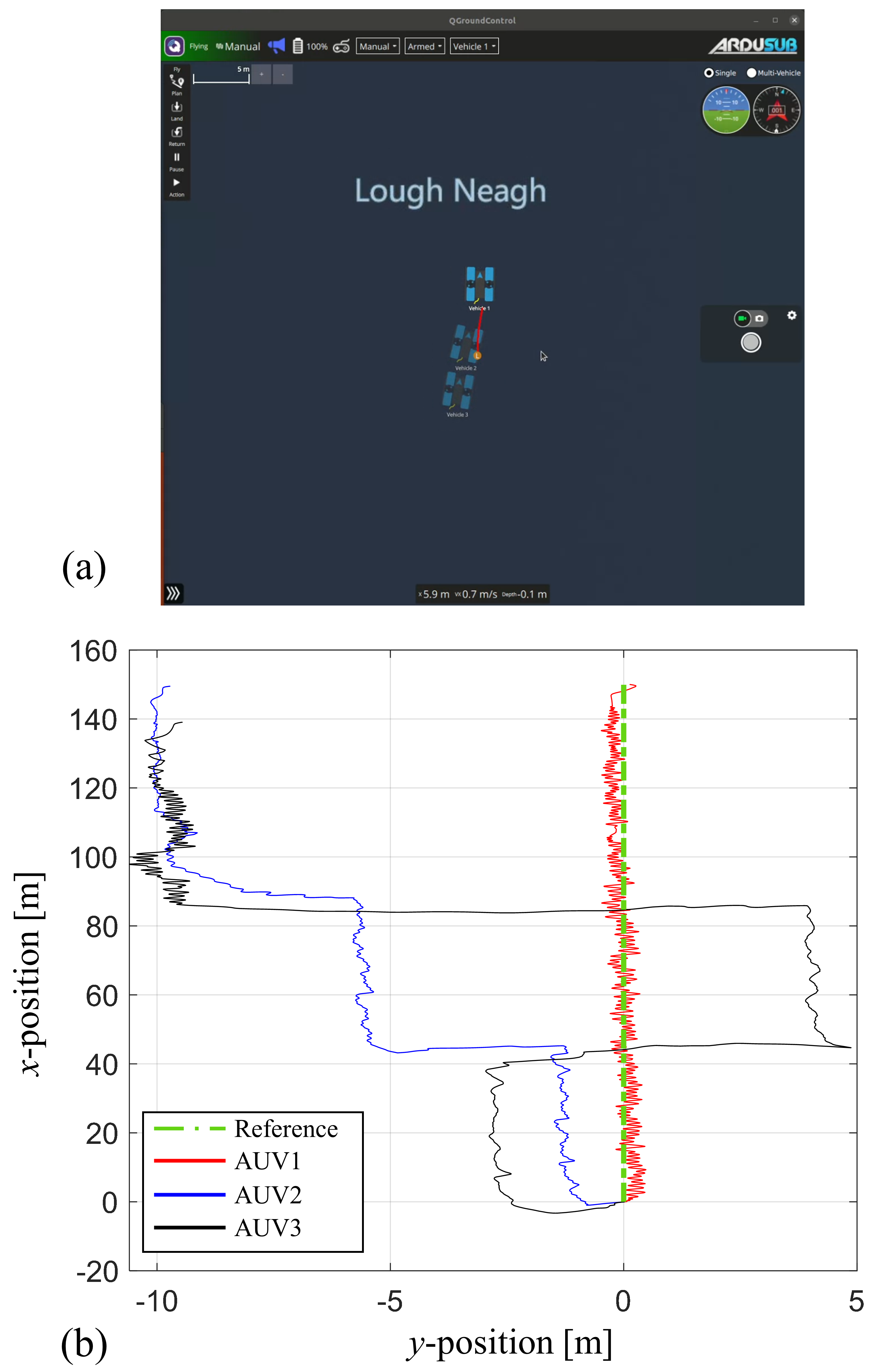}
\caption{Trajectory tracking and formation control in SITL simulation.
\\(See Video 1 at \href{https://youtu.be/5e0GfazS348}{https://youtu.be/5e0GfazS348})}
\label{ros}
\end{figure}
\section{Conclusion}
In summary, this study introduces a novel Distributed Nonlinear Model Predictive Control (DNMPC) framework for real-time reconfiguration and connectivity maintainance of a AUVs network, integrating Control Lyapunov Functions (CLF) and Control Barrier Functions (CBF). This approach has been validated through MATLAB and SITL simulations, demonstrating enhanced manoeuvrability, safety, and efficient collaborative behaviour in dynamic aquatic environments. The results underline the potential of DNMPC for improving underwater exploration and task execution, paving the way for advanced multi-AUV operations. Future endeavours will continue focusing on the implementation of these control strategies onto physical AUVs (i.e., BlueRov2) platforms, facilitating their application in real-world scenarios.
\addtolength{\textheight}{-12cm}   

\end{document}